# Inflation and CMB Anisotropy from Quantum Metric Fluctuations

Leonid Marochnik and Daniel Usikov
Physics Department, East-West Space Science Center, University of Maryland, College Park, MD 20742

**Abstract**

We propose a model of cosmological evolution of the early and late Universe which is consistent with observational data and naturally explains the origin of inflation and dark energy. We show that the de Sitter accelerated expansion of the FLRW space with no matter fields (hereinafter, empty space) is its natural state, and the model does not require either a scalar field or cosmological constant or any other hypotheses. This is due to the fact that the de Sitter state is an exact solution of the rigorous mathematically consistent equations of one-loop quantum gravity for the empty FLRW space that are finite off the mass shell. Space without matter fields is not empty as it always has the natural quantum fluctuations of the metric, i.e. gravitons. Therefore, the empty (in this sense) space is filled with gravitons, which have the backreaction effect on its evolution over time forming a self-consistent de Sitter instanton leading to the exponentially accelerated expansion of the Universe. At the start and the end of cosmological evolution, the Universe is assumed to be empty, which explains the origin of inflation and dark energy. This scenario leads to the prediction that the signs of the parameter 1+w should be opposite in both cases, and this fact is consistent with observations. The fluctuations of the number of gravitons lead to fluctuations of their energy density which in turn leads to the observed CMB temperature anisotropy of the order of 10^-5 and CMB polarization. In the frame of this scenario, it is not a hypothetical scalar field that generates inflation and relic gravitational waves but on the contrary, the gravitational waves (gravitons) generate dark energy, inflation, CMB anisotropy and polarization.

## 1. Introduction

As is known, the idea of a possible key role of de Sitter solution for the early stages of evolution of the Universe goes back to de Sitter, Eddington and Lemaitre. Gliner [1] was the first to show that the equation of state $p = -\varepsilon$, generating de Sitter expansion of the Fiedmannien Universe is the equation of state of anti-gravitating physical vacuum. On the basis of this idea, Zeldovich [2] showed that the cosmological constant producing the de Sitter expansion of the Friedmannien Universe is of the vacuum nature. In the pioneering work of Starobinsky [3], it was shown by direct computation that the quantum corrections to Einstein's equations that come from conformal anomalies lead to the nonsingular de Sitter solution instead of a standard Big Bang singularity (in some works, this solution is considered as an alternative to scalar field inflation, see e.g. [4]). The possibility of de Sitter expansion at the beginning of cosmological evolution was also found in works [5, 6]. Later on, it was turned out that the appearance of the de Sitter equation of state in the empty early Universe is not accidental because any quantum corrections to the Einstein equations lead to the energy density term of the form

$$\varepsilon = C \cdot \hbar H^4 \qquad (1.1)$$

Where $C$ is a dimensionless constant, $H$ is Hubble constant and $\hbar$ is Planck constant. Such a form of the energy density term is simply a consequence of its dimension [7, section V]. It is usual to call space without matter fields empty. Hereinafter, the term "empty space" is used in this sense. In fact, space without matter fields is not empty, as it always has the natural quantum fluctuations of the metric, i.e.

gravitons. Therefore, the empty (in this sense) space is filled with gravitons, which have the backreaction effect on its evolution over time. The quantum state of gravitons is determined by their interaction with a macroscopic field, and the macroscopic (background) geometry, in turn, depends on the state of gravitons. The background metric and the graviton operator appearing in the self–consistent theory are extracted from the unified gravitational field, which initially satisfies exact equations of quantum gravity. The classical component of the unified field is by definition a function of coordinates and time. The quantum component of the same unified field is described by a tensor operator function, which also depends on coordinates and time. Under such formulation of the problem, the original exact equations should be the operator equations of quantum theory of gravity in the Heisenberg representation. The derivation of these equations and their relation to existing references can be found in [8]. For the first time, these equations (and exact solutions) were given by [9]. In work [9], the exact equations of one-loop quantum gravity for the empty space with FLRW geometry were obtained in the form of Bogoliubov-Born-Green-Kirkwood-Yvon hierarchy or BBGKY- chain. One of the exact solutions to the equations of the chain turned out to be the de Sitter solution. Later on, this exact de Sitter solution was obtained directly from the rigorous mathematically consistent equations of one-loop quantum gravity in works [7], [8] and [10]. Technically, the exact de Sitter solution is a solution to these equations in imaginary time $\tau$. As was shown in work [10], the de Sitter solution is invariant with respect to Wick rotation $t = i\tau$ (see identity (24)), and due to this fact it can be applied to the real time Universe. It is important to note that the first rigorous mathematically consistent equations of one- loop quantum gravity obtained in works [7], [8], [9] and [10] *are finite off the graviton mass shell*[1]. In the empty space, they are the exact equations of self-consistent theory of gravitons in the Heisenberg representation with the ghost sector automatically providing one-loop finiteness off the mass shell. These equations are the only mathematically consistent of all in the available literature [8]. Because of conformal non–invariance and zero rest mass of gravitons, no conditions exist in the Universe to place gravitons on the mass shell precisely. Therefore, in the absence of one–loop finiteness, divergences arise in observables. To eliminate them, the Lagrangian of Einstein's theory must be modified, by amending the definition of gravitons. In other words, in the absence of one–loop finiteness, gravitons generate divergences, contrary to their own definition [7, 8]. Such a situation does not make any sense, so the one–loop finiteness off the mass shell is a prerequisite for internal consistency of the theory. As already noted, one of exact solutions to these equations of one- loop quantum gravity that are finite off the graviton mass shell is the de Sitter solution.

Thus, there is no need for the cosmological constant, hypothetical scalar fields and/or other hypotheses to generate the de Sitter exponential expansion of the empty FLWR Universe because such an expansion is its natural state.

**2. Thresholds**

At the first time, the Universe was empty (presumably) at the beginning of its cosmological evolution. The second time, it is going to became empty by the end of its cosmological evolution when the energy density of non-relativistic matter is approaching zero. Thus, one has to expect to find de Sitter expansion at the beginning and at the end of cosmological evolution of the Universe. At the beginning, it generates inflation, so that there is no need for hypothetical scalar fields. By the end, it generates the observed cosmological acceleration which is known as dark energy effect, and there is no need in the cosmological

---

[1] The one-loop quantum gravity with no matter fields is finite on the graviton mass shell [11]

constant or other hypotheses. If so, both the inflation and dark energy are one and the same macroscopic quantum effect of graviton condensation on the horizon scale of the non–stationary Universe. The instanton nature of this solution leads to the appearance of thresholds in both cases. As was already mentioned, the exact de Sitter solution was obtained in imaginary time for the empty space and then analytically continued into real time. In the presence of matter, we have to follow the same procedure because a continuous transition to the case of empty space should exist. In the presence of gravitons and matter, the first Friedmannien equation reads

$$3H^2 = 8\pi G(\varepsilon_g + \varepsilon_m) \qquad (2.1)$$

In the Euclidean space of imaginary time $\tau = -it$, it reads [10]

$$3H_\tau^2 = 8\pi G(|\varepsilon_{inst}| - \varepsilon_m) \qquad (2.2)$$

Where the energy density of instantons is $\varepsilon_{inst}$, the energy density of matter is $\varepsilon_m$ [10]. To be tunneled into the Lorentzian space of the Universe, solutions to (2.2) must exist. They do exist if the energy density of matter is below the threshold

$$\varepsilon_m \leq |\varepsilon_{inst}| \qquad (2.3)$$

In the dark energy case, the energy density of non-relativistic matter is initially higher than this threshold, i.e. $\varepsilon_m > |\varepsilon_{inst}|$. So, when $\varepsilon_m$ drops to the threshold (2.3) (and below) it marks the birth of dark energy which is a possible explanation to the fact that the birth of dark energy is taking place during the contemporary epoch of cosmological evolution ("coincidence problem") [10]. In the case of inflation, the situation is reversed in the following sense. Presumably, there was no matter initially, so the Universe started with the natural de Sitter expansion. With time, the new-born matter began to appear; so that de Sitter expansion was changed to quasi-de Sitter, and finally, the inflation stops when the energy density of a new-born matter $\varepsilon_M$ has increased up to the threshold $\varepsilon_M = |\varepsilon_{inst}|$. After that the standard Big Bang cosmology begins. It is convenient to characterize the difference between the cases of inflation and dark energy by the parameter $\delta = -\dot{H}/H^2$ (so-called "slow-roll" parameter in the scalar field theory) where the sign of $\delta$ indicates the direction of change in $H$. If the Hubble function $H$ is increasing with time then $\delta < 0$ and vice versa. Note that $\delta = \delta_\tau$ where $\delta_\tau = -\dot{H}_\tau / H_\tau^2$ (here the dot is derivative over $\tau$). In the case of DE, one has to expect an increase in $H_\tau$ with time because $\varepsilon_m \sim a^{-3}$ is decreasing (see Eq. (2.2)), and hence $\delta_{DE} < 0$. In the case of inflation, the energy density of a new-born matter $\varepsilon_M$ is increasing with time (until it stops after it reaches the threshold (2.3)), so one has to expect $\delta_{inf} > 0$ in this case. Such behavior of parameter $\delta$ is consistent with observations (see section 4).

The following example is to illustrate the inflation case. The energy density of particles into which gravitons decayed satisfies a conservation equation [13]

$$\dot{\varepsilon}_M + 3H(\varepsilon_M + p_M) = \Gamma \varepsilon_g e^{-\Gamma(t-t_s)} \qquad (2.4)$$

The right-hand side of (2.4) takes into account the flow of energy from gravitons. Here $\varepsilon_g$ is energy density of gravitons forming de Sitter solution, $\Gamma$ is the rate of decay of gravitons into other particles, and $t_{start} \equiv t_s$ is taken at the beginning of the process of filling the empty space with the new-born matter.

Assuming that products of decay of gravitons are highly relativistic, i.e. $\varepsilon_M = 3p_M$, we find the following solution to the Eq. (2.4)

$$\varepsilon_M = a^{-4} \int_{t_s}^{t} \Gamma \varepsilon_g a^4 e^{-\Gamma(t'-t_s)} dt' \tag{2.5}$$

The energy density of new-born matter (2.5) starts equal to zero at $t = t_s$ and then first rises up to some maximum value and finally falls as the density is attenuated by the expansion of the Universe. The inflation is supposed to be stopped when the energy density of the new-born matter $\varepsilon_M$ reaches the threshold $\varepsilon_M = |\varepsilon_{inst}|$. We do not know how primordial gravitons decay into other particles but we can consider some hypothetical example for pure illustrative purposes. In accordance with the inflation paradigm, at a sufficiently early stage of cosmological evolution all modes of interest must be deep inside the horizon which means that we deal with short wave gravitons. As is known [14], in this case[2]

$$\varepsilon_g = \varepsilon_g^{\,0}(a_s/a)^4 \tag{2.6}$$

Taking into account Eq. (2.6), we get from (2.5)

$$\varepsilon_M = \varepsilon_g^{\,0}(\frac{a_s}{a})^4 [1 - e^{-\Gamma(t-t_s)}] \tag{2.7}$$

One can see from (2.7) that at $t = t_s$ the energy density of the new-born matter $\varepsilon_M(t_s) = 0$, then it increases to the maximum $\varepsilon_M^{\,max} / \varepsilon_g^{\,max} = (1 + 4H/\Gamma)^{-1}$ where $\varepsilon_g^{\,max} = \varepsilon_g^{\,0}(a_s/a_{max})^4$, and after that the expansion continues with the usual Big Bang factor $\varepsilon_M \sim a^{-4}$. The inflation continues until the energy density of a new-born matter is increasing up to the threshold $\varepsilon_M = |\varepsilon_{inst}|$. After that it stops, and the standard Big Bang cosmology begins. This means that for the above example the condition $\varepsilon_M^{\,max} \leq |\varepsilon_{inst}|$ must be satisfied. The birth of DE is considered in our work [10]. The existence of threshold in the DE case allows a natural explanation of the "coincidence problem". In short, the solutions to Eq. (2.2) can exist only when the density of non-relativistic matter drops below the threshold (2.3) which corresponds to the redshift $1+z_T = (\Omega_{DE}/\Omega_m)^{1/3}$, and this threshold is consistent with observational data [10].

---

[2] Eq. (2.6) is valid for the gravitons of short wavelengths, i.e. for the condition $k\eta \gg 1$ where the conformal time is $\eta = \int dt/a$. In the process of expanding, the wavelengths of gravitons are stretched out, and after crossing the horizon they satisfy the reverse condition $k\eta \ll 1$ which produces the long wave modes for which instead of (2.6) we get $\varepsilon \sim a^{-2}$ and $\varepsilon \sim a^{-6}$ [15] (see also [7], section IV B).

## 3. CMB anisotropy and polarization from fluctuations of the number of gravitons

The equation of state of graviton condensate forming de Sitter solution described above reads [7, 9] (cf. (1.1))

$$p = -\varepsilon = -\frac{3N\hbar H^4}{8\pi^2} \quad (3.1)$$

Here $N$ is a dimensionless functional of parameters of state vector which is of the order of the number of gravitons in the Universe expanding with the Hubble velocity $H$. According to [7] (section VII), in such a condensate the phases in the graviton and ghost sectors correlate similarly, and the non-zero effect appears due to the difference of average occupation numbers for graviton's and ghost's instantons. In the Universe, the graviton condensate produced by quantum tunneling provided by instantons. The equation of state (3.1) is superficially similar to that which comes from quantum conformal anomalies [3]. Quantum corrections to the Einstein equations due to zero oscillations can provide a self-consistent De Sitter solution in the vicinity of Planck's value of curvature ([3], [16], [17]). In such a case, the equation of state is $\varepsilon \sim n\hbar H^4$ where the number of types of elementary particles $n$ is of the order of $\leq 100$ [18]. Conformal anomalies that arise due to regularization and renormalization procedures do not apply to the graviton condensate which is the solution of the equations of quantum gravity that are finite in the one-loop approximation. In the finite one-loop quantum gravity, the effect of conformal anomalies is exactly zero, and the De Sitter solution can be formed only by graviton-ghost instantons ([7], section X). In contrast to the conformal anomaly parameter $n$, the parameter $N$ is arbitrary and can be a huge number.

In our case, it is of the order of the number of gravitons in the Universe $N_g$, i.e. $N = \alpha N_g$ where $\alpha = O(1)$. In the contemporary Universe, we have $N_g \sim 10^{122}$ [9, 10].

The equation of state (3.1) was obtained under the assumption that the spectrum is flat and typical occupation numbers in the ensemble are large, so that squares of modules of probability amplitudes are likely to be described by Poisson distribution [7]. If so, from Eq. (3.1) it follows that in the energy units $\hbar = c = 1$ $H$ reads

$$H^2 = 8\pi^2 \frac{M_{pl}^2}{\alpha N_g} \quad (3.2)$$

Where $M_{pl}$ is the reduced Planck mass $M_{pl} = (\hbar c/8\pi G)^{1/2} = 2.4 \cdot 10^{18} GeV/c^2$. From (3.2) it follows that the fluctuations of the number of gravitons $N_g$ that are presumably described by Gaussian distribution are $<(\Delta N_g)^2>/<N_g>^2 = <N_g>^{-1}$, so that

$$\frac{<(\Delta N_g)^2>}{<N_g>^2} = \frac{\alpha}{8\pi^2} \cdot \frac{H^2}{M_{pl}^2} \qquad (3.3)$$

Here $<(\Delta N_g)^2> = <N_g^2> - <N_g>^2$. Due to Eq. (3.1), one gets

$$\frac{<(\Delta N_g)^2>}{<N_g>^2} = \frac{<(\Delta \varepsilon)^2>}{<\varepsilon>^2} \qquad (3.4)$$

Thus, fluctuations of the number of gravitons produce fluctuations of energy density. They play the same role as scalar perturbations (density fluctuations) which are responsible for the anisotropy of CMB in the models of inflation using scalar fields. In other words, if the graviton condensate is responsible for the inflation then fluctuations of number of gravitons are the cause of the anisotropy of CMB. For the typical energy scale of inflation $H \simeq 10^{15} GeV$ and $\alpha = 1$ one gets from (3.3) and (3.4)

$$(\frac{<(\Delta \varepsilon)^2>}{<\varepsilon>^2})^{1/2} \simeq 1.5 \cdot 10^{-5} \qquad (3.5)$$

As is known, temperature fluctuations $\Delta T / T$ are of the same order of magnitude as the metric and density perturbations which contribute directly to $\Delta T / T$ via the Sachs-Wolfe effect. Thus, the fluctuations of the number of gravitons in the Universe are able to produce CMB anisotropy $\Delta T / T \sim 10^{-5}$ due to fluctuations of gravitational potential which in turn are of the order of fluctuations of energy density. Any mechanism generating the temperature anisotropy inevitably generates the CMB polarization as well, which is an order of magnitude below the temperature fluctuations ([19] and references therein). The B-mode of polarization of CMD is produced by gravitational waves. Over the de Sitter background, the mode functions of gravitons are (see, e.g. [9])

$$f(k\eta) = (1 - \frac{i}{k\eta})e^{-ik\eta} \qquad f^+(k\eta) = (1 + \frac{i}{k\eta})e^{ik\eta} \qquad (3.6)$$

The Bunch-Davies initial conditions select $f(k\eta)$ as a unique mode function in theories of inflation based on the scalar field hypothesis. Gravitons of instanton origin select the same $f(k\eta)$ as a unique mode function but because of a different reason[3], so that there is no difference between power spectra of tensor perturbations produced by scalar fields and gravitons produced by instantons. Thus, both temperature fluctuations and B-mode of polarization of CMB can be of the same origin. Both of them can be produced by primordial gravitons.

## 4. Spectrum of metric fluctuations

---

[3] Instanton solutions are sought in imaginary time by transition $\eta \to -i\eta$. Requirement of finiteness eliminates the $f^+(k\eta)$ solution (see [7] section VII and [11]), so only $f(k\eta)$ is the a unique mode function in this case

The de Sitter solution considered above is produced by the flat spectrum of metric fluctuations [7, 10]. The spectrum slightly deviating from the flat spectrum produces the quasi-de Sitter solution. The observed tilt $n_s - 1$ of power spectrum $k^{n_s - 1}$ deviates slightly from the scale-invariant form corresponding to $n_s = 1$. The observed value is $n_s \approx 0.96$ [20]. This means that in reality we deal with a quasi-de Sitter expansion. To consider this case, we use Eq. (15) from [10] which can be presented in the following form

$$H^2 = \frac{6\pi^2 M_{pl}^2}{\int_0^\infty N_k [-\xi^2 + (1+\xi)^2] e^{-2\xi} \xi \, d\xi} \tag{4.1}$$

In the case of a flat spectrum, $N_k = N = \alpha N_g = const$, and (4.1) leads to (3.2) (after calculation of the integral in the denominator). Assuming that $N_k = N_0 (k/k_0)^\beta$ where $k_0$ is a pivot scale, we get from (4.1)

$$H^2 = \frac{6\pi^2 M_{pl}^2 (k_0 \eta)^\beta}{N_0 \int_0^\infty [-\xi^2 + (1+\xi)^2] e^{-2\xi} \xi^{1+\beta} \, d\xi} \tag{4.2}$$

The straightforward calculation $\delta$ from (15) leads to $\delta = \beta/2$. From Friedmannien equations it follows that $\delta = 3(1+w)/2$ (where $w = p/\varepsilon$ is the equation of state parameter), so we get

$$\beta = 3(1+w) \tag{4.3}$$

In the present approach, the deviations from de Sitter solution are due to any kind of matter that makes the space is non-empty. So, the parameter $\beta$ is the one that characterizes the degree of by how much the space is filled by matter. In case of DE, the Planck data [21] obtained by combination of Planck+WP+BAO give, e.g. $w = -1.13_{-0.25}^{+0.24}$, i.e. $\beta_{DE} = 2\delta_{DE} \approx -0.39 < 0$, and the energy density of non-relativistic mater which is close to $\Omega_m \approx 0.3$. One can say that $\beta_{DE} \approx -0.39$ corresponds to the Universe, ~ 70 percent of which is already empty. Note the negative sign of $\beta_{DE}$ as was expected for the dark energy case. Assuming homogeneous and isotropic Gaussian processes, we get the dimensionless variance for the two-point correlation function of fluctuations of the number of gravitons

$$\frac{<\delta N_k \delta N_{k'}>}{N_k^2} = \frac{k^3}{2\pi^2} \frac{1}{N_k} = \frac{k^3}{2\pi^2} \frac{1}{N_0} \left(\frac{k}{k_0}\right)^{-\beta} \tag{4.4}$$

From (4.4) it follows that $n_s - 1 = -\beta$, and for $n_s \approx 0.96$, we get for the inflation case $\beta_{inf} = 2\delta_{inf} \approx 0.04 > 0$. Note the positive sign of $\beta_{inf}$ as was expected for the inflation case. Thus, the signs of parameter $\beta = 2\delta$ are consistent with the instanton origin of DE and inflation.

## 6. Conclusion

Based on the above consideration, the following scenario of cosmological evolution of the early and late Universe can be proposed. The de Sitter expansion is a natural state of empty FLRW space, so there is no need in scalar fields, cosmological constant and other hypotheses to generate such an expansion of the empty space. After the (quantum) birth of the Universe, it was presumably empty and the natural de Sitter expansion generates the inflation. In the process of filling the Universe with a new-born matter, de Sitter expansion becomes first quasi-de Sitter and then it stops when the energy density of the new-born matter reaches the threshold (energy density of instantons). After that the standard Big Bang cosmology begins. Nearing the end of the evolution, the Universe empties, and its expansion is approaching again the de Sitter law what is observed as the dark energy effect. This scenario is consistent with observational data. Fluctuations of the number of gravitons generate the CMB anisotropy of the order of $\Delta T / T \sim 10^{-5}$ and inevitably they must generate the CMB polarization; the difference in signs of "slow-roll" parameter $\delta$ in cases of inflation and dark energy is consistent with the prediction of instanton theory; the threshold after which the dark energy appears is also consistent with observations (coincidence problem) [10].

**Acknowledgment**

We are grateful to Arthur Chernin for graciously agreeing to read our manuscript and make comments on our approach. One of us (LM) is grateful to Natalia Ptitsyna for stimulating discussions. We would like to express our deep appreciation and special thanks to Walter Sadowski for invaluable advice and help in the preparation of the manuscript.